\documentclass[a4paper,10pt]{article}
\usepackage[utf8]{inputenc}
\usepackage{amsmath}
\usepackage{amsfonts}
\usepackage{float}

\title{$SU(4)$-Ward identities for QCD with restored chiral symmetry}
\author{
Vasily Sazonov \footnote{\texttt{vasily.sazonov@th.u-psud.fr}}\\
Institute of Physics, University of Graz,\\ Universit\"atsplatz 5, A-8010 Graz, Austria\\
Laboratoire de Physique Th\'eorique, CNRS UMR 8627,\\ 
Universit\'e Paris XI,  F-91405 Orsay Cedex, France}

\begin{document}
 
\maketitle

\begin{abstract}
Lattice studies of QCD at temperatures above the chiral restoration and QCD
with truncated low modes of the Dirac operator indicate
approximate and explicit $SU(4)$ degeneracies in hadron spectra, respectively.
At the same time, the QCD classical action and the path integral measure are not invariant under $SU(4)$.
Here we investigate $SU(4)$ transformations in the continuum limit by deriving corresponding Ward identities.
We show that, if there is a gap in low-lying modes of the Dirac operator the
obtained $SU(4)$ Ward identities are simplified and look like they would be 
if $SU(4)$ symmetry is preserved. Then, we discuss possible consequences for the quark matter at high temperatures.

\end{abstract}

\section{Introduction}
The notion of symmetry plays a fundamental role in theoretical physics.
In classical mechanics, continuous symmetries necessarily lead
to the conservation of Noether's currents. 
In quantum physics, symmetries of classical action can be preserved or spontaneously and/or anomalously broken.
An interesting question is what happens after the quantization with transformations, which
are not symmetries at the classical level, {\it i.e.}
do not leave the classical action invariant. {\it A priori} one can not exclude the possibility
that quantization can act in opposite to the standard intuition and may turn such transformations into the
emergent (or in other words effective) symmetries of the system. 

One of the most convenient and powerful approaches to study symmetries and transformations in quantum field theory
is the framework of Ward identities. 
These identities describe an invariance of the path integral with respect to
an arbitrary infinitesimal change of variables (and can be also derived for transformations, which are not symmetries of the classical action). 
Therefore, we propose to utilize the possible simplicity or complexity of Ward identities as a criteria for symmetries. 
Namely, if a Ward identity under certain conditions reduces to
a Noether's conservation law, then we say that the system is symmetric with respect to the corresponding transformation.
Obviously, such definition unifies emergent (effective) and standard symmetries.

In the present paper we apply such general ideas to Quantum Chromodynamics. In a series of works by L. Glozman and collaborators
\cite{Lang2011, Glozman2012, Denissenya2014, Glozman2014, Denissenya20142, Markus2015, Denissenya2015, Denissenya20152}
the relation between zero modes of the QCD Dirac operator and chiral symmetry breaking (Banks-Casher relation \cite{Banks1980}) was investigated numerically.
It was shown there, that an artificial removing
of the low modes of the Dirac operator in lattice computations leads to the vanishing of the chiral condensate
and to the degeneracy in hadron spectra. However, the degeneracy in spectra contained much more
particles, than it was required by the restoration of chiral symmetry. 
It was suggested in \cite{Glozman2014} and precisely shown in \cite{Markus2015} that correlators
of particles with degenerate masses are connected by the $SU(4)$ transformation.
Recently, a lattice study of QCD at temperatures above the chiral phase transition \cite{Rohrhofer2017}
pointed out approximate $SU(4)$ degeneracies of hadron correlators.


In the following we investigate the emergence of the $SU(4)$ symmetry within the framework
of the Ward identities and mentioned above definition of effective symmetries.
In Section \ref{SU2def} we present basic definitions of $SU(4)$. In Section \ref{WI} we derive
$SU(4)$ Ward identities. By studying the spectral decomposition of Ward identities in Section \ref{spectrd},
we show that removing of the zero and near zero modes drastically simplifies $SU(4)$ Ward identities,
turning them into the Noether conservation laws. 
In Section \ref{rest} we conclude and discuss possible relations between the chiral restoration, confinement and and the emergence of $SU(4)$ symmetry.

\section{$SU(4)$ transformations}
\label{SU2def}
The quark part of the QCD Lagrangian with $N_F = 2$ flavors with degenerate masses in the Euclidean space can be written as
\begin{equation}
{\cal L} = \overline\Psi (\mathbf{1}_F \otimes (\gamma_\mu {D}_\mu + m)) \Psi\,,~~~~~~~~\mu = 1..4\,,
\label{L}
\end{equation}
where $\Psi$ stands for a vector containing both quark flavors,
the $\gamma_\mu$ matrices are hermitian and obey 
$\gamma_\mu\gamma_\nu + \gamma_\mu\gamma_\nu = 2 \delta_{\mu\nu}$ and $\gamma_\mu{D}_\mu = \gamma_\mu(\partial_\mu - i g A_\mu)$ 
is the euclidean Dirac operator in the background gluon field $A_\mu = A_\mu^a t^a$ with $t^a$ being generators of the $SU(N)$ gauge group.
The transformations of the $SU(4)$ group are completely determined by generators $T_l$, 
satisfying the ${\mathbf{su}}(4)$ algebra commutation relations
\begin{eqnarray}
\nonumber
  [T^l, T^m] = 2 i f^{lmn}T^n\,,~~~ f^{lmn} = \frac{1}{8i}\text{Tr}[[T^l, T^m]T^n]\,,\\
  \{T^l, T^m\} = 2\delta^{lm} (\mathbf{1}_F\otimes\mathbf{1}_D) + 2 d^{lmn} T^n\,,~~~d^{lmn} = \frac{1}{8}\text{Tr}[\{T^l, T^m\}T^n]\,,
\end{eqnarray}
where $f^{lmn}$ denotes fully antisymmetric structure constants, 
$d^{lmn}$ is a fully symmetric tensor, $l,m,n=1,2,..,15$ and $\mathbf{1}_F$, $\mathbf{1}_D$ are the unit matrices in flavor and Dirac spaces, correspondingly.
The generators $T^l$ are given by $15$ matrices
\begin{equation}
  T = \{(\tau^a \otimes \mathbf{1}_D), (g_F \otimes \Sigma^i)\}\,,~~~g_F = \{1_F, \tau^a\}
\label{Tl}
\end{equation}
with indices $a, i = 1,2,3$.
The Pauli matrices $\tau^a$ in (\ref{Tl}) are the generators of the standard $SU(2)_L \times SU(2)_R$ chiral symmetry.
The matrices $\Sigma^i$ are
\begin{equation}
  \Sigma^i = \{\gamma_4,\, i\gamma_5\gamma_4,\,-\gamma_5\}
\end{equation}
and satisfy the ${\mathbf{su}}(2)$ algebra commutation relations
\begin{equation}
  [\Sigma^i,\Sigma^j] = 2 i \epsilon^{ijk} \Sigma^k\,.
\end{equation}
The generators $\Sigma^i$ define chiral spin $SU(2)_{CS}$ subgroup of $SU(4)$ with
transformations acting only in the Dirac space
\begin{equation}
  \Psi \rightarrow \Psi' = e^{i\mathbf{\epsilon(x)} \cdot (\mathbf{1}_F\otimes \Sigma)}\Psi\,.
\label{trpsi}
\end{equation}
The transformations of the full  $SU(4)$ group 
\begin{equation}
  \Psi \rightarrow \Psi' = e^{i\mathbf{\epsilon(x)} \cdot T}\Psi\,,~~~
  \Psi = 
    \left(\begin{array}{c}
      u_L\\
      u_R\\
      d_L\\
      d_R
     \end{array}\right)
\label{PsiT}
\end{equation}
mix both quark flavors and left-/right-handed components.

In the Euclidean space spinors $\Psi$ and $\overline{\Psi}$ are completely independent.
Thus, one can consider their independent transformations.
The transformations, which belong also to the chiral symmetry group,
must act on the spinor $\overline{\Psi}$ in the same way as they usually defined in the Euclidean space
(for instance, being compatible with the axial anomaly).
Therefore, we define the transformation of the spinor $\overline{\Psi}$, using $\gamma_0 = -i\gamma_4$, as follows.
Let $\widetilde\Psi = -i\Psi^\dagger\gamma_4$, then its transformations with respect to $SU(4)$
are determined by (\ref{PsiT})
\begin{equation}
  \widetilde\Psi \rightarrow \widetilde\Psi' = -i(e^{i\mathbf{\epsilon(x)} \cdot T}\Psi)^\dagger\gamma_4\,.
\label{tildPsi}
\end{equation}
The action of $SU(4)$ on the spinor $\overline{\Psi}$ is defined 
to be equivalent of the $SU(4)$ action on the $\widetilde\Psi$,
we write it as
\begin{equation}
  \overline\Psi \rightarrow \overline\Psi' = \overline\Psi\, e^{i\mathbf{\epsilon(x)} \cdot \overline{T}}\,,~~~
  \overline{T} = \{(-\tau^a \otimes \mathbf{1}_D),(g_F\otimes \overline\Sigma^i)\}\,.
\label{PsiBT}
\end{equation}
For the infinitesimal $\epsilon(x)$ the variations of $\Psi$ and $\overline\Psi$
are $\delta \Psi = i\epsilon(x) T \Psi$ and $\delta \overline\Psi = i\epsilon(x) \overline{T}\, \overline\Psi$.
The $\delta\Psi$ and $\delta\overline{\Psi}$ for 15 generators of $SU(4)$, satisfying the rule (\ref{tildPsi}), 
are presented in the Table \ref{TablGen}.
%
\begin{table}[H]
\begin{center}
\begin{tabular}{l|l|l}
Generators & $\delta\Psi$ & $\delta\overline{\Psi}$\\
\hline
$\mathbf{1}_F\otimes\gamma_4$ & $i \epsilon(x) \gamma_4 \Psi$ & $-i\epsilon(x) \overline\Psi \gamma_4$\\
$\mathbf{1}_F\otimes i\gamma_5\gamma_4$ & $-\epsilon(x) \gamma_5 \gamma_4 \Psi$ & $-\epsilon(x) \overline\Psi \gamma_5 \gamma_4$\\
$\mathbf{1}_F\otimes(-\gamma_5)$ & $-i \epsilon(x)\gamma_5\Psi$ & $-i \epsilon(x) \overline\Psi \gamma_5$\\
$\tau_a\otimes\mathbf{1}_D$ & $i \epsilon(x)\tau_a\Psi$ & $-i \epsilon(x) \overline\Psi \tau_a$\\
$\tau_a\otimes\gamma_4$ & $i \epsilon(x) \tau_a \gamma_4 \Psi$ & $-i\epsilon(x) \overline\Psi \gamma_4 \tau_a$\\
$\tau_a\otimes i\gamma_5\gamma_4$ & $-\epsilon(x) \tau_a \gamma_5 \gamma_4 \Psi$ & $-\epsilon(x) \overline\Psi \gamma_5 \gamma_4 \tau_a$\\
$\tau_a\otimes(-\gamma_5)$ & $-i \epsilon(x) \tau_a \gamma_5\Psi$ & $-i \epsilon(x) \overline\Psi \gamma_5 \tau_a$
\end{tabular}
\end{center}
\caption{Infinitesimal variations of spinors $\Psi$ and $\overline\Psi$ under the $SU(4)$ transformations,
in the last two columns of the table the tensorial product of flavor and Dirac structures is omitted to shorter notations.}
\label{TablGen}
\end{table}

\section{Ward identities}
\label{WI}
\subsection{Lagrangian transformations}
\label{LT}
The quark Lagrangian (\ref{L}) is not invariant under the $SU(4)$ transformations.
In the current section we present its variations depending on generators (\ref{Tl}), 
see Table \ref{TableLagr}. 
\begin{table}[H]
\begin{center}
\begin{tabular}[!htb]{l|l}
Generators & $\delta{\cal L} = \delta\overline\Psi\, {\cal D}\, \Psi + \overline\Psi\, {\cal D}\, \delta\Psi$\\
\hline
$\mathbf{1}_F\otimes\gamma_4$ & 
  $i(\partial_\mu \epsilon(x))\overline\Psi\gamma_\mu\gamma_4\Psi -$
  $2 i \epsilon(x) \overline{\Psi}\gamma_4 \gamma_i[\partial_i - i g A_i]\Psi$\\

$\mathbf{1}_F\otimes i\gamma_5\gamma_4$ & 
  $-(\partial_\mu \epsilon(x)) \overline\Psi \gamma_\mu\gamma_5\gamma_4\Psi -
  2 m \epsilon(x) \overline{\Psi}\gamma_5\gamma_4\Psi $\\~&$
  -2 \epsilon(x)\overline{\Psi}\gamma_i\gamma_5\gamma_4 [\partial_i - i g A_i]\Psi$\\
  
$\mathbf{1}_F\otimes(-\gamma_5)$ & 
  $-i(\partial_\mu \epsilon(x)) \overline\Psi \gamma_\mu\gamma_5\Psi - 2 i\epsilon(x) m \overline{\Psi}\gamma_5\Psi$\\

$\tau_a\otimes\mathbf{1}_D$ & 
  $i(\partial_\mu \epsilon(x)) \overline\Psi \gamma_\mu\tau_a\Psi$\\
  
$\tau_a\otimes\gamma_4$ & 
  $i(\partial_\mu \epsilon(x))\overline\Psi\gamma_\mu\gamma_4\tau_a\Psi - 
  2 i \epsilon(x) \overline{\Psi}\gamma_4 \gamma_i \tau_a [\partial_i - i g A_i]\Psi$\\

$\tau_a\otimes i\gamma_5\gamma_4$ & 
  $-(\partial_\mu \epsilon(x)) \overline\Psi \gamma_\mu\gamma_5\gamma_4\tau_a\Psi-
  2 m \epsilon(x) \overline{\Psi}\gamma_5\gamma_4\tau_a\Psi $\\~&$
  -2 \epsilon(x)\overline{\Psi}\gamma_i\gamma_5\gamma_4 \tau_a [\partial_i - i g A_i]\Psi$\\

$\tau_a\otimes(-\gamma_5)$ & 
  $(\partial_\mu \epsilon(x)) \overline\Psi \gamma_\mu\gamma_5 \tau_a \Psi - 2 i\epsilon(x) m \overline{\Psi}\gamma_5 \tau_a\Psi$
\end{tabular}
\end{center}
\caption{Variations of the quark Lagrangian under the $SU(4)$ transformations.
Here ${\cal D} = (\mathbf{1}_F \otimes (\gamma_\mu {D}_\mu + m))$ 
and in the right part of the table the tensorial product of flavor and Dirac structures is omitted to shorter notations.}
\label{TableLagr}
\end{table}

\subsection{Measure transformation}
The non-invariance of the path integral measure with respect to a certain transformation usually is referred
as an anomalous violation of symmetry. Here we keep this name regarding $SU(4)$ transformations.
To study corresponding properties of the spinor measure, we follow \cite{Fujikawa1979} 
(see also \cite{bookNakahara} for a derivation with hermitian matrices). 
We consider QCD in a finite volume with
periodic boundary conditions for gluons in all directions and with anti-periodic/periodic boundary conditions
for quarks in temporal/spacial directions.
Then, the Dirac operator $i\gamma_\mu{D}_\mu$, being hermitian, has a real discrete spectrum and complete set of orthonormal
eigenvectors:
\begin{eqnarray}
\nonumber
  i\gamma_\mu{D}_\mu \phi_n(x) = \lambda_n \phi_n(x)\,,~~~\lambda_n \in \mathbb{R}\,,\\
\nonumber
  \int d^4 x\, \phi^\dagger_{m}(x) \phi_{n}(x) = \delta_{m, n}\,,\\
  \sum_n \phi_{n}(x) \phi^\dagger_n(y) = \delta(x - y)\,.
\label{spectrum}
\end{eqnarray}
Using the basis of the Dirac operator eigenfunctions, two independent two flavor spinors $\Psi$ and $\overline\Psi$
can be represented as
\begin{eqnarray}
\nonumber
  \Psi(x) = 
      \left(\begin{array}{c}
	\sum_n a_n^{(1)} \phi_n(x)\\
	\sum_n a_n^{(2)} \phi_n(x)
     \end{array}\right)\\
\nonumber
\\
  \overline\Psi(x) = 
      \left(\begin{array}{c}
	\sum_m \phi^\dagger_m(x) \overline b_m^{(1)}\\
	 \sum_m \phi^\dagger_m(x) \overline b_m^{(2)}
     \end{array}\right)\,,
\end{eqnarray}
where indices $(1)$ and $(2)$ label explicitly the first and the second quark flavor.
Then, the measure of the path integral can be written as
\begin{equation}
  d\mu \equiv \prod_x [D A_\mu(x)] \prod_{f = 1}^{2}\prod_{n, m} d\overline{b}_m^{(f)} da_n^{(f)}\,.
\label{measure}
\end{equation}
The change of the measure (\ref{measure}) caused by the application
of $SU(4)$ transformations (\ref{PsiT}), (\ref{PsiBT}) is
\begin{equation}
    \Psi(x) \rightarrow \Psi'(x) \equiv e^{i \epsilon(x) g_F \otimes \Sigma} \Psi(x) \equiv 
      \left(\begin{array}{c}
	\sum_n \tilde{a}_n^{(1)} \phi_n(x)\\
	\sum_n \tilde{a}_n^{(2)} \phi_n(x)
     \end{array}\right)\\
\end{equation}
where
\begin{equation}
  \tilde{a}_n^{(f)} = \sum_{\hat f}\sum_m \int d^4 x\, \phi^\dagger_n\,  e^{i \epsilon(x) (g_F \otimes \Sigma)}\, \phi_m(x) a_m^{(\hat f)} 
  = \sum_{\hat f} \sum_m C^{f, \hat f} \otimes C_{n,m} a_m^{(\hat f)}\,.
\end{equation}
Then,
\begin{equation}
  \prod_n da_n = \big[\det (C^{f, \hat f} \otimes C_{k,l})\big]^{-1}\prod_n da'_n\,,
\end{equation}
\begin{eqnarray}
\nonumber
  \big[\det (C^{f, \hat f} \otimes C_{k,l})\big]^{-1} = \det[\delta_{k,l} + i \int d^4x \epsilon(x) \phi^\dagger_k(x)(g_F \otimes  \Sigma)\phi_l(x)]^{-1} = \\
  \exp[-i \text{Tr}_F[g_F] \int d^4x \epsilon(x) \sum_k \phi^\dagger_k(x)\,\Sigma\, \phi_k(x)]\,,
\end{eqnarray}
where we have explicitly written the trace in the flavor space. 
The transformation of the measure $\prod_m d\overline{b}_m$ 
is determined by rules (\ref{PsiBT})
\begin{equation}
  \prod_m d\overline{b}_m = \big[\det (\overline C^{f, \hat f} \otimes \overline C_{k,l})\big]^{-1}\prod_m d\overline{b}'_m\,,
\end{equation}
\begin{equation}
  \big[\det (\overline C^{f, \hat f} \otimes \overline C_{k,l})\big]^{-1} = 
  \exp[-i\text{Tr}_F[g_F] \int d^4x \epsilon(x) \sum_k \phi^\dagger_k(x)\,\overline{\Sigma}\,\phi_k(x)]\,.
\end{equation}
Therefore, the anomaly term of the action is given by
\begin{equation}
  -i\text{Tr}_F[g_F] \int d^4x \epsilon(x) \sum_k \phi^\dagger_k(x)\Big[\Sigma + \overline{\Sigma}\Big]\phi_k(x)\,.
\label{sumSigm}
\end{equation}
When $g_F = \mathbf{1}_F$, $\text{Tr}_F[g_F] = 2$, alternatively, one has $\text{Tr}_F[g_F = \tau_a] = 0$. Consequently, there is no
anomaly for all generators of $SU(4)$, containing $\tau_a$. The anomalous violation is also absent 
for generators $(g_F\otimes\gamma_4)$ due to the cancellation of $\Sigma$ and $\overline\Sigma$. 
For the generator $(\mathbf{1}_F\otimes(i\gamma_5\gamma_4))$, without explicit computations, the anomaly
term can be expressed as
\begin{equation}
  4 i \int d^4x \epsilon(x) \lim_{M\rightarrow\infty} \sum_k \phi^\dagger_k(x) \gamma_5\gamma_4 e^{-(\lambda_k / M)^2} \phi_k(x)\,,
\end{equation}
where the regularization $e^{-(\lambda_k / M)^2}$ is introduced, to assign the meaning to the sum over the eigenmodes.

\subsection{Identities}
To derive $SU(4)$-Ward identities, we combine together the information from
Table \ref{TableLagr} and computations of the anomalous contributions.
We treat terms containing $\partial_\mu \epsilon(x)$ employing periodic boundary
conditions for $\epsilon(x)$ and periodic/anti-periodic boundary conditions for quark fields.
The bilinear forms containing the mixing of quark flavors performed by $\tau^a$
vanish after the integration over the quark fields and one ends up with the three following identities
\begin{eqnarray}
\nonumber
\label{ltWI1}
    &&\partial_\mu\langle\overline\Psi\gamma_\mu\gamma_4\Psi\rangle_A + 
    2 \langle\overline{\Psi}\gamma_4 \gamma_i[\partial_i - i g A_i]\Psi\rangle_A = 0\,,\\
\nonumber
    &&\partial_\mu \langle\overline\Psi \gamma_\mu\gamma_5\gamma_4\Psi\rangle_A
    -2 m \langle \overline{\Psi}\gamma_5\gamma_4\Psi\rangle_A
    -2 \langle\overline{\Psi}\gamma_i\gamma_5\gamma_4 [\partial_i - i g A_i]\Psi\rangle_A \\
\nonumber
    &&+4 i \lim_{M\rightarrow\infty} \sum_k \phi^\dagger_k(x) \gamma_5\gamma_4 e^{-(\lambda_k / M)^2} \phi_k(x) = 0\,,\\
\nonumber
    &&-\partial_\mu \langle\overline\Psi \gamma_\mu\gamma_5\Psi\rangle_A - 2 i m \langle \overline{\Psi}\gamma_5\Psi\rangle_A
    -\frac{1}{4\pi^2}\,\text{Tr}\,^*F_{\mu\nu}F_{\mu\nu} = 0\,,\\
\end{eqnarray}
where $^*F_{\mu\nu}$ is the dual field-strength tensor, brackets $\langle..\rangle_A$ stand for the 
averaging over quark fields at some fixed gluon configuration and eigenfunctions $\phi_k(x)$ of the one-flavor Dirac operator
also depend on the same fixed gluon configuration.


\section{Spectral properties of $SU(4)$-identities}
\label{spectrd}
\subsection{Classical part of identities}
The spectral content of the terms of $SU(4)$-Ward identities can be studied in the following way. 
In the most general form the terms, related to the non-invariance
of the classical action, look like
$\langle\overline\Psi_{\alpha, a}^{(f)}(x) O_{\alpha\beta}^{ab} \Psi_{\beta, b}^{(f)}(x)\rangle_A$
with a flavor index $f$, $\alpha$, $\beta$ and $a$, $b$ being Dirac and color indices respectively. 
Hereafter we omit all indices, unless it is necessary for understanding.
Integrating over the spinor fields and utilizing
the spectral representation, we obtain
\begin{equation}
  \langle\overline\Psi(x) O \Psi(x)\rangle_A = 2 \sum_n \phi^\dagger_n(x) O \phi_n(x) \frac{1}{m - i\lambda_n}\,,
\label{sp1}
\end{equation}
where $\phi_n(x)^\dagger$ and $\phi_n(x)$ are eigenfunctions of the one-flavor Dirac operator, as before,
and $2$ in front of the sum over $n$ comes from the flavor trace. 
The summation over $n$ in (\ref{sp1}) is ill-defined and has to be regularized, precisely as in the previous section 
(and as in the derivation of the axial anomaly)
\begin{equation}
  \langle\overline\Psi(x) O \Psi(x)\rangle_A = 2 \lim_{M\rightarrow\infty} 
  \sum_n \phi^\dagger_n(x) O e^{-(\lambda_n / M)^2} \phi_n(x) \frac{1}{m - i\lambda_n}\,.
\label{sp2}
\end{equation}
Then, we split the sum in (\ref{sp2}) into the three parts: with positive, negative and zero eigenvalues
\begin{eqnarray}
\nonumber
  \langle\overline\Psi(x) O \Psi(x)\rangle_A = \\ 
  2 \lim_{M\rightarrow\infty} 
  \Big(\sum_{\lambda_n > 0} + \sum_{\lambda_n < 0} + \sum_{\lambda_n = 0}\Big) \phi^\dagger_n(x) O e^{-(\lambda_n / M)^2} \phi_n(x) \frac{1}{m - i\lambda}\,.
\label{sp3}
\end{eqnarray}
The anti-commutation of $\gamma_5$ with matrices $\gamma_\mu$ ensures that eigenfunctions, 
corresponding to non-zero eigenvalues, come in pairs 
\begin{equation}
  i \gamma_\mu D_\mu (\gamma_5\phi_n) = -\lambda_n (\gamma_5\phi_n) \equiv \lambda_{-n} \phi_{-n}
\label{g51}
\end{equation}
and
\begin{equation}
  \phi_{-n} = \gamma_5\phi_n~~~\Rightarrow~~~\phi_{-n}^\dagger = \phi_{n}^\dagger \gamma_5\,.
\label{g52}
\end{equation}
Now we join the sums over the positive and negative eigenvalues
\begin{eqnarray}
\nonumber
  \langle\overline\Psi(x) O \Psi(x)\rangle_A = 2 \lim_{M\rightarrow\infty} \Bigg[
  \sum_{\lambda_n = 0} \phi^\dagger_n(x) O \phi_n(x) \frac{1}{m} + \\
\nonumber
  \sum_{\lambda_n > 0} 
  \Big(
  \phi^\dagger_n(x) O e^{-(\lambda_n / M)^2} \phi_n(x) \frac{1}{m - i\lambda_n}\\
  +
  \phi^\dagger_n(x)\gamma_5 O e^{-(\lambda_n / M)^2} \gamma_5 \phi_n(x) \frac{1}{m + i\lambda_n}
  \Big)\Bigg]\,.
\label{sp4}
\end{eqnarray}
Since both operators $O = \{\gamma_4 \gamma_i[\partial_i - i g A_i], \gamma_i\gamma_5\gamma_4 [\partial_i - i g A_i]\}$ commute with $\gamma_5$,
$[O, \gamma_5] = 0$,
two terms of the last sum in (\ref{sp4}) can be easily grouped together
\begin{eqnarray}
\nonumber
  \langle\overline\Psi(x) O \Psi(x)\rangle_A = 2 \lim_{M\rightarrow\infty} \Bigg[
  \sum_{\lambda_n = 0} \phi^\dagger_n(x) O \phi_n(x) \frac{1}{m} + \\
  \sum_{\lambda_n > 0} 
  \phi^\dagger_n(x) O e^{-(\lambda_n / M)^2} \phi_n(x) \frac{2\, m}{\lambda_n^2 + m^2}
  \Bigg]\,.
\label{sp5}
\end{eqnarray}
Then, it is evident that in the zero mass limit the dominant contributions to the expectation values
$\langle\overline\Psi(x) O \Psi(x)\rangle_A$ come from zero modes of the Dirac operator.

\subsection{Anomalous part} 
In identities (\ref{ltWI1}) the term responsible for the anomalous breaking $U(1)_A$ symmetry is equal 
to the difference in the amount of right-handed and left-handed zero modes of the Dirac operator.
Consequently, it is equal to zero, when there is a gap in the spectral density.

Another term in (\ref{ltWI1}) produced by the non-invariance of the measure is associated with the 
$(\mathbf{1}_F\otimes(i\gamma_5\gamma_4))$ generator of $SU(4)$.
To study it, we first perform regularization 
\begin{equation}
  4 i \langle\sum_k\phi_k^\dagger\gamma_5\gamma_4\phi_k\rangle_A = \lim_{M\rightarrow\infty}
  4 i \langle\sum_k\phi_k^\dagger\gamma_5\gamma_4 e^{-(\lambda_k / M)^2} \phi_k\rangle_A\,.
\label{g4g5an}
\end{equation}
Taking into account (\ref{g52})
the sum over $k$ in (\ref{g4g5an}) can be decomposed into the sum over zero modes ($k = 0$) and
modes with positive eigenvalues ($k > 0$)
\begin{eqnarray}
\nonumber
  &&\sum_k \phi^\dagger_k(x)\gamma_5\gamma_4 e^{-(\lambda_k / M)^2} \phi_k(x) = 
  \sum_{k = 0} \phi^\dagger_k(x)\gamma_5\gamma_4 e^{-(\lambda_k / M)^2} \phi_k(x) +\\
\nonumber
  &&\sum_{k > 0} \Big(\phi^\dagger_k(x)\gamma_5\gamma_4 e^{-(\lambda_k / M)^2} \phi_k(x) + 
  \phi^\dagger_{-k}(x)\gamma_5\gamma_4 e^{-(\lambda_k / M)^2} \phi_{-k}(x) \Big) = \\
\nonumber
  &&\sum_{k = 0}\phi^\dagger_k(x)\gamma_5\gamma_4 e^{-(\lambda_k / M)^2}\phi_k(x) + 
  \sum_{k > 0} \phi^\dagger_k(x)[\gamma_5,\gamma_4] e^{-(\lambda_k / M)^2} \phi_k(x) = \\
  &&\sum_{k = 0}\phi^\dagger_k(x)\gamma_5\gamma_4 e^{-(\lambda_k / M)^2} \phi_k(x)\,.
\end{eqnarray}
Thus, the anomalous term (\ref{g4g5an}) is also built 
only from the zero eigenmodes of the Dirac operator and it is zero if there is a spectral gap.

\section{Concluding remarks: possible\\ restoration of $SU(4)$}
\label{rest}
Finally, in the zero mass limit, 
a gap at the origin in the Dirac spectrum causes the reduction of 
the $SU(4)$ Ward identities (\ref{ltWI1}) to a set of conservation laws
\begin{eqnarray}
    \partial_\mu\langle\overline\Psi\gamma_\mu\gamma_4\Psi\rangle_A = 0\,,~~~
    \partial_\mu \langle\overline\Psi \gamma_\mu\gamma_5\gamma_4\Psi\rangle_A = 0\,,~~~
    \partial_\mu \langle\overline\Psi \gamma_\mu\gamma_5\Psi\rangle_A= 0\,.
\label{ltWIRes}
\end{eqnarray}
The latter equations provide an analytical explanation of the emergence of $SU(4)$ degeneracies in lattice simulations
with truncated low modes of the Dirac operator \cite{Glozman2012, Glozman2014} 
\footnote{The simplification of the identities (\ref{ltWI1}) gives only a partial explanation, 
since a full one requires a study of spectral properties of all possible Ward identities associated with $SU(4)$ transformations.}.

The possible emergence of the $SU(4)$ symmetry in the reality is directly related to the eventual appearance
of a gap in the Dirac spectrum under certain conditions, which still remains an open question.
The most recent lattice simulations with manifestly chirally-invariant fermions \cite{Bazavov2012, Cossu2013, Tomiya2017} show the generation
of a spectral gap at the temperatures above the chiral restoration ($T > T_c$).
However, such predictions typically strongly depend on the lattice volumes utilized in computations,
and for the numerical proof of the possible existence of the spectral gap additional studies
on the bigger lattices are needed. 

From the point of view of analytical derivations,  it is known, that in accordance with 
the Banks-Casher relation \cite{Banks1980}, the chiral condensate is proportional to the density of the 
zero modes of the Dirac operator. Since, the chiral condensate vanishes for temperatures $T > T_c$,
the density of zero modes also must vanish, $\rho(\lambda = 0) = 0$.
The most precise estimate proven by nowadays for lattice-regularized QCD with overlap fermions \cite{Aoki2012}
gives $\rho(\lambda) \leq const \cdot \lambda^\alpha$, $\alpha > 2$, when $\lambda\rightarrow 0$,
what does not forbid an existence of the spectral gap, but also does not prove it.
Therefore, if the gap opens at some temperature $T_{c_2} \geq T_c$, then the $SU(4)$ symmetry gets effectively
restored at this temperature and provides an example of the quantum emergent symmetry, which does not leave invariant
the classical action.

If the gap does not open, then it is possible at list one more scenario.
The $U(1)_A$ is a subgroup of $SU(4)$ and the emergence of $SU(4)$ is closely related to the effective restoration of $U(1)_A$.
According to previous studies \cite{Cohen1996, Cohen1998} and current analysis the existence of the spectral gap is a sufficient condition 
for the effective restoration of $U(1)_A$.
At the same time, axial anomaly is absent in odd spacial dimensions \cite{bookNakahara} and, consequently, must not exist in the infinite
temperature limit $T \rightarrow \infty$ due to the compactification of time.
Then, it is natural to assume, that the limit $T \rightarrow \infty$ in some sense is equivalent to the presence of the gap in the Dirac spectrum,
and that the distribution of the Dirac eigenmodes approaches the distribution containing a gap with the temperature increase. 
If such scenario indeed realizes in Nature, the $SU(4)$ symmetry has to emerge gradually, as $T \rightarrow \infty$.

\subsection*{Acknowledgments}
I am grateful to Leonid Glozman and Markus Pak for numerous discussions and for the introduction to the topic of the $SU(4)$ symmetry.
I would like to thank Christian Lang for the careful reading of the manuscript and for pointing several drawbacks.
Support from the Austrian Science Fund
(FWF) through the grant P26627-N27 and the Erwin Schrödinger fellowship J-3981 is acknowledged.

\label{Bibliography}

\end{document}